\documentclass[journal]{IEEEtran}
\usepackage{fullpage,graphicx,psfrag,epsfig,amsmath,amsfonts,amssymb,fancybox}
\usepackage[dvips]{color}
\begin{document}

\title{An Outer Bound to the Capacity Region of the
Broadcast Channel}

\author{Chandra~Nair,~\IEEEmembership{Member,~IEEE,}
        and~Abbas~El~Gamal,~\IEEEmembership{Fellow,~IEEE.}
\thanks{Chandra Nair is a post-doctoral researcher with the Theory Group at Microsoft Research.}
\thanks{Abbas El Gamal is a professor with the Electrical
Engineering Dept. at Stanford University.}}

\markboth{IEEE }{An Outer Bound to the Capacity Region of the
Broadcast Channel}

\maketitle

\newcommand{\eq}{\begin{equation}}
\newcommand{\en}{\end{equation}}
\newcommand{\eqa}{\begin{eqnarray}}
\newcommand{\ena}{\end{eqnarray}}
\newcommand{\eqas}{\begin{eqnarray*}}
\newcommand{\enas}{\end{eqnarray*}}
\newcommand{\ra}{ {\rightarrow} }
\newcommand{\Ra}{ {\Rightarrow} }
\newcommand{\bX}{\mathbf{X}}
\newcommand{\bY}{\mathbf{Y}}
\newcommand{\bZ}{\mathbf{Z}}
\newcommand{\bU}{\mathbf{U}}
\newcommand{\bV}{\mathbf{V}}
\newcommand{\bW}{\mathbf{W}}
\newcommand{\p}{{\rm P}}
\def\mb#1{\mathbf{#1}}
\newenvironment{myproof}[1]{\noindent\hspace{2em}{\itshape Proof #1:}}{\hspace*{\fill}~\par\endtrivlist\unskip}
%calligraphic letters%
\def\cA{{\mathcal A}}
\def\cB{{\mathcal B}}
\def\cC{{\mathcal C}}
\def\cD{{\mathcal D}}
\def\cE{{\mathcal E}}
\def\cF{{\mathcal F}}
\def\cG{{\mathcal G}}
\def\cH{{\mathcal H}}
\def\cI{{\mathcal I}}
\def\cJ{{\mathcal J}}
\def\cK{{\mathcal K}}
\def\cL{{\mathcal L}}
\def\cM{{\mathcal M}}
\def\cN{{\mathcal N}}
\def\cO{{\mathcal O}}
\def\cP{{\mathcal P}}
\def\cQ{{\mathcal Q}}
\def\cR{{\mathcal R}}
\def\cS{{\mathcal S}}
\def\cT{{\mathcal T}}
\def\cU{{\mathcal U}}
\def\cV{{\mathcal V}}
\def\cW{{\mathcal W}}
\def\cX{{\mathcal X}}
\def\cY{{\mathcal Y}}
\def\cZ{{\mathcal Z}}

%defining colors

\definecolor{Red}{rgb}{1,0,0}
\definecolor{Blue}{rgb}{0,0,1}
\definecolor{Green}{rgb}{0,1,0}
\definecolor{Yellow}{rgb}{1,1,0}
\definecolor{Cyan}{rgb}{0,1,1}
\definecolor{Magenta}{rgb}{1,0,1}
\definecolor{Orange}{rgb}{1,.5,0}
\definecolor{Violet}{rgb}{.5,0,.5}
\definecolor{Purple}{rgb}{.75,0,.25}
\definecolor{Brown}{rgb}{.75,.5,.25}
\definecolor{Grey}{rgb}{.5,.5,.5}

\def\red{\color{Red}}
\def\blue{\color{Blue}}
\def\green{\color{Green}}
\def\yellow{\color{Yellow}}
\def\cyan{\color{Cyan}}
\def\magenta{\color{Magenta}}
\def\orange{\color{Orange}}
\def\violet{\color{Violet}}
\def\purple{\color{Purple}}
\def\brown{\color{Brown}}
\def\grey{\color{Grey}}

\def\tU{\tilde{U}}
\def\tV{\tilde{V}}
\def\tX{\tilde{X}}
\def\tW{\tilde{W}}
\def\tY{\tilde{Y}}
\def\tZ{\tilde{Z}}

\def\cc{\bar{\mbox{co}}}

\newtheorem{theorem}{Theorem}[section]
\newtheorem{lemma}[theorem]{Lemma}
\newtheorem{claim}[theorem]{Claim}
\newtheorem{conjecture}[theorem]{Conjecture}
\newtheorem{corollary}[theorem]{Corollary}
\newtheorem{definition}[theorem]{Definition}
\newtheorem{example}[theorem]{Example}
\newtheorem{xca}[theorem]{Exercise}
\newtheorem{fact}[theorem]{Fact}
\newtheorem{remark}[theorem]{Remark}

\renewcommand \thesection{\arabic{section}}

\newcommand{\enp} {\hfill \rule{2.2mm}{2.6mm}}
\def\mb#1{\mathbf{#1}}
\numberwithin{equation}{section}

%    Absolute value notation
%\newcommand{\abs}[1]{\lvert#1\rvert}

%    Blank box placeholder for figures (to avoid requiring any
%    particular graphics capabilities for printing this document).

\newcommand{\blankbox}[2]{%
  \parbox{\columnwidth}{\centering
%    Set fboxsep to 0 so that the actual size of the box will match the
%    given measurements more closely.
    \setlength{\fboxsep}{0pt}%
    \fbox{\raisebox{0pt}[#2]{\hspace{#1}}}%
  }%
}

\begin{abstract}

An outer bound to the capacity region of the two-receiver discrete
memoryless broadcast channel is given. The outer bound is tight
for all cases where the capacity region is known. When specialized
to the case of no common information, this outer bound is
contained in the K\"{o}rner-Marton outer bound. This containment
is shown to be strict for the {\em binary skew-symmetric}
broadcast channel. Thus, this outer bound is in general tighter
than all other known outer bounds.

\end{abstract}

%\IEEEpeerreviewmaketitle

\section{Introduction}

We consider a discrete memoryless (DM) broadcast channel where the sender
wishes to communicate common as well as separate messages to two
receivers~\cite{cov72}. Formally, the channel consists of an input alphabet
$\mathcal{X}$, output alphabets $\mathcal{Y}$ and $\mathcal{Z}$, and
a probability transition function $p(y,z|x)$.  A
$((2^{nR_0},2^{nR_1}, 2^{nR_2}),n)$ code for this channel consists
of (i) three messages $(M_0,M_1,M_2)$ uniformly distributed over
$[1,2^{nR_0}]\times [1, 2^{nR_1}]\times [1,2^{nR_2}]$, (ii) an
encoder that assigns a codeword $x^n(m_0,m_1,m_2)$, for each message
triplet $(m_0,m_1,m_2) \in [1,2^{nR_0}]\times [1, 2^{nR_1}]\times
[1,2^{nR_2}]$, and (iii) two decoders, one that maps each received
$y^n$ sequence into an estimate $(\hat m_0,\hat m_1)\in
[1,2^{nR_0}]\times [1, 2^{nR_1}]$ and another that maps each
received $z^n$ sequence into an estimate $(\hat{\hat{m_0}},\hat
m_2)\in [1,2^{nR_0}]\times [1, 2^{nR_2}]$.

The probability of error is defined as
\begin{align*}
\label{eq:perrorcominfo}
 P_e^{(n)} &= \p(\hat M_0\neq M_0 \; \mbox{\rm or } \hat{\hat{M_0}}\neq
 M_0 \; \mbox{\rm or } \hat M_1\neq M_1 \\
 & \qquad ~~ \mbox{\rm or } \hat M_2\neq M_2).
\end{align*}
A rate tuple $(R_0, R_1,R_2)$ is said to be achievable if there exists
a sequence of $((2^{nR_0}, 2^{nR_1}, 2^{nR_2}),n)$ codes with
$P_e^{(n)} \rightarrow 0$. The capacity region of the broadcast
channel is the closure of the set of
achievable rates.

The capacity region for this channel is known only for some
classes, including the degraded \cite{ber73,gal74,ahk75}, less
noisy\cite{kom75}, more capable\cite{elg79},
deterministic\cite{mar77,pin78} and semi-deterministic
channels\cite{gep80}. Additionally, general inner bounds by Cover
\cite{cov75}, van der Meulen~\cite{van75} and Marton~\cite{mar79}
and outer bounds by K\"{o}rner and Marton~\cite{mar79} and
Sato~\cite{sat78} have been established. Furthermore, the
K\"{o}rner and Marton~\cite{mar79} outer bound was found to be
tight for all cases where capacity is known.

In this paper we introduce an outer bound on the capacity region
of the DM broadcast channel based on results in~\cite{elg79} and
show that it is strictly tighter than existing outer bounds. The
outer bound is presented in the next section. In
Section~\ref{se:nocommon}, the outer bound is specialized to the
case of no common information. In Section~\ref{subse:km}, it is
shown that that when there is no common information, our outer
bound is contained in the K\"{o}rner-Marton bound and in
Section~\ref{se:bssc} it is shown that this containment is strict.

\section{Outer bound}
\label{se:obound}
The following is an outer bound to the capacity region of the
two-receiver DM broadcast channel.

\begin{theorem}
\label{th:outer}  The set of rate triples $(R_0, R_1, R_2)$  satisfying
\begin{equation}
\label{eq:obound0}
\begin{aligned}
  R_0 & \leq \min \{ I(W;Y), I(W;Z) \}, \\
  R_0 + R_1 & \leq I(U,W;Y), \\
  R_0 + R_2 & \leq I(V,W;Z),  \\
  R_0 + R_1 + R_2 & \leq  I(U,W;Y) + I(V;Z|U,W), \\
  R_0 + R_1 + R_2 & \leq I(V,W;Z) + I(U;Y|V,W), \notag \\
  \end{aligned}
\end{equation}
for some joint distribution of the form $p(u,v,w,x) = p(u)p(v)p(w|u,v)p(x|u,v,w)$ constitutes an outer bound to the capacity region for the DM
broadcast channel.
\end{theorem}

\begin{proof} The arguments are essentially the same as those used in the
converse proof for the more capable broadcast channel class \cite{elg79}.
Observe that
\begin{align*}
nR_0 &= H(M_0) \\
&= H(M_0|Y^n) + I(M_0;Y^n) \\
&\stackrel{(a)}{\leq} n\lambda_{0n} + \sum_{i=1}^n
\left(H(Y_i|Y^{i-1}) - H(Y_i|M_0,Y^{i-1})\right) \\
&\stackrel{(b)}{\leq} n\lambda_{0n} + \sum_{i=1}^n \left(H(Y_i) -
H(Y_i|M_0,Y^{i-1}, Z_{i+1}^n)\right),
\end{align*}
where (a) follows by Fano's inequality and (b) follows from the
fact that conditioning decreases entropy. Now defining the random
variable $W_i = (M_0,Y^{i-1}, Z_{i+1}^n)$, we obtain
\begin{equation}
\label{eq:r0y}
\begin{aligned}
nR_0 & \leq n\lambda_{0n} + \sum_{i=1}^n \left(H(Y_i) - H(Y_i|W_i)\right)\\
& = n\lambda_{0n} + \sum_{i=1}^n I(Y_i;W_i).
\end{aligned}
\end{equation}
In a similar fashion observe that
\begin{equation}
\begin{aligned}
\label{eq:r0z}
nR_0 & = H(M_0) \\
& = H(M_0|Z^n) + I(M_0;Z^n) \\
& \leq n\lambda_{1n} + \sum_{i=1}^n
\left(H(Z_i|Z_{i+1}^n) - H(Z_i|M_0,Z_{i+1}^n)\right) \\
& \leq n\lambda_{1n} + \sum_{i=1}^n \left(H(Z_i) -
H(Z_i|M_0,Y^{i-1}, Z_{i+1}^n)\right) \\
& = n\lambda_{1n} + \sum_{i=1}^n I(Z_i;W_i).
\end{aligned}
\end{equation}
Now, consider
\begin{equation}
\label{eq:r1y}
\begin{aligned}
& n(R_0 + R_1)\\
&\quad = H(M_0,M_1) \\
&\quad = H(M_0,M_1|Y^n) + I(M_0,M_1; Y^n) \\
&\quad \leq n\lambda_{2n} + \sum_{i=1}^n
\left(H(Y_i|Y^{i-1}) - H(Y_i|M_0,M_1, Y^{i-1})\right) \\
&\quad \leq  n\lambda_{2n} + \sum_{i=1}^n
\left(H(Y_i) - H(Y_i|M_0,M_1, Y^{i-1},Z_{i+1}^n)\right)\\
&\quad \leq n\lambda_{2n} + \sum_{i=1}^n I(Y_i;U_i,W_i),
\end{aligned}
\end{equation}
where we define the random variable $U_i= M_1$ for all $i$.

In a similar fashion
\begin{equation}
\label{eq:r2z} n(R_0 + R_2) \leq n\lambda_{3n} + \sum_{i=1}^n
I(Z_i;V_i,W_i),
\end{equation}
where $V_i=M_2$ for all $i$.

Lastly, consider
\begin{equation*}
\begin{aligned}
&n(R_0 + R_1 + R_2)\\
&\quad = H(M_0,M_1,M_2) \\
&\quad = H(M_0,M_1) + H(M_2|M_0,M_1) \\
&\quad \leq n\lambda_{4n} + I(M_0,M_1;Y^n) +
I(M_2;Z^n|M_0,M_1) \\
\end{aligned}
\end{equation*}
\begin{equation}
\label{eq:r1r2ypre}
\begin{aligned}
&\quad = n\lambda_{4n} + \sum_{i=1}^n I(M_0,M_1;Y_i|Y^{i-1})\\
&\qquad \qquad ~ + \sum_{i=1}^n I(M_2;Z_i|M_0,M_1,Z_{i+1}^n).
\end{aligned}
\end{equation}
Note that
\begin{equation}
\label{eq:r1r2yint1}
\begin{aligned}
& \sum_{i=1}^n I(M_0,M_1;Y_i|Y^{i-1})\\
&\quad \leq \sum_{i=1}^n I(M_0,M_1,Y^{i-1};Y_i)\\
&\quad = \sum_{i=1}^n I(M_0,M_1,Y^{i-1},Z_{i+1}^n;Y_i) \\
&\quad  \qquad - \sum_{i=1}^n I(Z_{i+1}^n;Y_i|M_0,M_1,Y^{i-1}).
\end{aligned}
\end{equation}
And further,
\begin{equation} \label{eq:r1r2yint2}
\begin{aligned}
& \sum_{i=1}^n I(M_2;Z_i|M_0,M_1,Z_{i+1}^n)\\
&\quad \leq
\sum_{i=1}^n I(M_2,Y^{i-1};Z_i|M_0,M_1,Z_{i+1}^n)\\
&\quad =\sum_{i=1}^n I(Y^{i-1};Z_i|M_0,M_1,Z_{i+1}^n)\\
&\qquad +\sum_{i=1}^n I(M_2;Z_i|M_0,M_1,Z_{i+1}^n,Y^{i-1}).
\end{aligned}
\end{equation}

Combining equations \eqref{eq:r1r2ypre}, \eqref{eq:r1r2yint1},
\eqref{eq:r1r2yint2}, we obtain
\begin{equation*}
\begin{aligned}
& n(R_0 + R_1 + R_2)\\
 &\quad \leq n\lambda_{4n} + \sum_{i=1}^n
I(M_0,M_1,Y^{i-1},Z_{i+1}^n;Y_i)\\
&\qquad - \sum_{i=1}^n I(Z_{i+1}^n;Y_i|M_0,M_1,Y^{i-1}) \\
& \qquad + \sum_{i=1}^n I(Y^{i-1};Z_i|M_0,M_1,Z_{i+1}^n) \\
&\qquad + \sum_{i=1}^n I(M_2;Z_i|M_0,M_1,Z_{i+1}^n,Y^{i-1})
\end{aligned}
\end{equation*}
\begin{equation}
\label{eq:r1r2y}
\begin{aligned}
 &\quad \stackrel{(d)}{=} n\lambda_{4n}+
\sum_{i=1}^n
I(M_0,M_1,Y^{i-1},Z_{i+1}^n;Y_i)  \\
&\qquad + \sum_{i=1}^n I(M_2;Z_i|M_0,M_1,Z_{i+1}^n,Y^{i-1}) \\
&\quad = n\lambda_{4n} + \sum_{i=1}^n \left(I(U_i,W_i;Y_i) +
I(V_i;Z_i|U_i,W_i)\right).
\end{aligned}
\end{equation}
The equality (d) follows from the well known Csiszar identity (see
Lemma 7 in \cite{czk78}):
\[ \sum_{i=1}^n I(\bY^{i-1};Z_i|\bZ_{i+1}^n) =
\sum_{i=1}^n I(\bZ_{i+1}^n;Y_i|\bY^{i-1}). \]

In a very similar fashion, we can also obtain
\begin{equation}
\label{eq:r1r2z}
\begin{aligned}
&n(R_0 + R_1 + R_2)\\
&\quad \leq n\lambda_{5n} + \sum_{i=1}^n \left(I(V_i,W_i;Z_i) +
I(U_i;Y_i|V_i,W_i)\right).
\end{aligned}
\end{equation}

Define the time sharing random variable $Q$ to be independent of
$M_0,M_1,M_2,X^n,Y^n,Z^n,$ and uniformly distributed over $\{1,2,..,n\}$
and define $W=(Q,W_Q),\, U=U_Q,\, V=V_Q,\, X=X_Q,\, Y=Y_Q,\, Z=Z_Q$.

Clearly we have
\begin{equation}
\label{eq:regionproof1}
\begin{aligned}
nR_0 &\leq n\lambda_{0n} + \sum_{i=1}^n I(W_i;Y_i) \\
& = n\lambda_{0n} + n I(W;Y|Q) \\
& \leq n\lambda_{0n} + n I(W;Y). \\
\end{aligned}
\end{equation}
Similarly,
\begin{equation}
\label{eq:regionproof2}
\begin{aligned}
nR_0 &\leq n\lambda_{1n} + n I(W;Z) \\
n(R_0 + R_1) & \leq n\lambda_{2n} + n I(U,W;Y)\\
n(R_0 + R_2) & \leq n\lambda_{3n} + n I(V,W;Z).
\end{aligned}
\end{equation}
Further,
\begin{equation}
\label{eq:regionproof3}
\begin{aligned}
&n(R_0+R_1+R_2)\\
&\quad \leq n\lambda_{4n} + \sum_{i=1}^n
\left(I(U_i,W_i;Y_i) +
I(V_i;Z_i|U_i,W_i)\right) \\
& \quad = n\lambda_{4n} + n I(U,W;Y|Q) + nI(V;Z|U,W,Q) \\
& \quad = n\lambda_{4n}+ n I(U,W;Y|Q) + nI(V;Z|U,W) \\
& \quad \leq n\lambda_{4n}+ n I(U,W;Y) + nI(V;Z|U,W),
\end{aligned}
\end{equation}
and similarly
\begin{equation*}
\begin{aligned}
&n(R_0+R_1+R_2)\\
&\quad \leq n\lambda_{5n}+ n I(V,W;Z) + nI(U;Y|V,W).
\end{aligned}
\end{equation*}

The independence of the messages $M_1$ and $M_2$ implies the
independence of the auxiliary random variables $U$ and $V$ as
specified.

Since the probability of error is assumed to tend to zero, $
\lambda_{0n}, \lambda_{1n}, \lambda_{2n}, \lambda_{3n},
\lambda_{4n}$, and $\lambda_{5n}$ also tend to zero as $n\to \infty$. This completes the proof.

\end{proof}

\section{Outer Bound with No Common Information}
\label{se:nocommon}

Note that the outer bound given in Theorem~\ref{th:outer}
immediately leads to the following outer bound for the case when
there is no common information, i.e., $R_0 = 0$.

The set of all rate pairs $(R_1,R_2)$ satisfying
\begin{equation}
\label{eq:obound}
\begin{aligned}
  R_1 & \leq I(U,W;Y), \\
  R_2 & \leq I(V,W;Z),  \\
  R_1 + R_2 & \leq  I(U,W;Y) + I(V;Z|U,W)\\
  R_1 + R_2 & \leq I(V,W;Z) + I(U;Y|V,W), \\
\end{aligned}
\end{equation}
for some joint distribution of the form $p(u,v,w,x) =
p(u)p(v)p(w|u,v)p(x|u,v,w)$ constitutes an outer bound on the capacity of the
DM broadcast channel with no common information.

The following theorem gives a {\em possibly} weaker outer bound that we
consider for the rest of the paper.

\begin{theorem}
\label{th:obproper}
Consider the DM broadcast channel with no common information. The set of rate pairs $(R_1,R_2)$ satisfying
\begin{equation}
\begin{aligned}
  R_1 & \leq I(U;Y), \\
  R_2 & \leq I(V;Z),  \\
  R_1 + R_2 &  \leq  I(U;Y) + I(V;Z|U), \\
  R_1 + R_2 & \leq I(V;Z) +  I(U;Y|V), \notag \\
\end{aligned}
\end{equation}
for some choice of joint distributions
$p(u,v,x) = p(u,v)p(x|u,v)$ constitutes an outer bound to the capacity
region for the DM broadcast channel with no common information.
\end{theorem}

\begin{proof}
This follows by redefining $U$ as $(U,W)$ and $V$ as $(V,W)$ in
equation \eqref{eq:obound}.
\end{proof}

In the following subsections we prove  results that aid in the
evaluation of the above outer bound.

\subsection{$X$ Deterministic Function of $U,V$ suffices}

Denote by $\cC$ the outer bound in Theorem~\ref{th:obproper}
and let $\cC_d$ be the same bound but with $X$ restricted to be
a deterministic function of $U,V$, i.e. $\p(X=x|U=u,V=v)\in \{0,1\}$ for all $(u,v,x)$. We now show that these two bounds are identical.

\begin{lemma}
\label{le:bisuffices} $\cC = \cC_d$.
\end{lemma}

Before we prove this lemma, note that it suffices to show that
$\cC \subset \cC_d$.  Our method of proof is as follows: For every
$p(u,v), p(x|u,v)$ we will construct random variables
$U^*,V^*,X^*$ where $X^*$ is a deterministic function of $U^*,V^*$
such that the region described by $U^*,V^*,X^*$ will contain the
region described by $U,V,W$.

Let $\p(U=u,V=v) = p_{uv}$ and $\p(X=x|U=u,V=v) = \delta^x_{uv}\in
\{0,1\}$. Without loss of generality, assume that $\mathcal{X} =
\{0,1,..,m-1\}$. Now, construct random variables $U^*,V^*$ having
cardinalities $m\|U\|,m\|V\|$ as follows: Split each value $u$
taken by $U$ into $m$ values $u_0,\ldots,u_{m-1}$ and each value
$v$ taken by $V$ into $m$ values $v_0,\ldots,v_{m-1}$. Let
\begin{equation}
\label{eq:constr}
\begin{aligned}
&\p(U^*=u_i,V^*=v_j)\\
&\quad \quad = \frac{1}{m}\p(U=u,V=v,X=(i-j)_m), \\
&\p(X^*=k|U^*=u_i,V^*=v_j)\\
& \quad \quad = \left\{
\begin{array}{ll} 1 &\mbox{if } k=(i-j)_m\\[2pt] 0 &\mbox{otherwise,}
\end{array} \right.
\end{aligned}
\end{equation}
where $(l)_m$ is the remainder of $l/m$ (the $\mod$ operation).

We will need the following facts.
\begin{lemma} \label{le:construct}
The following hold:
\begin{itemize}
\item[(i)] $\p(U^*=u_i) = \frac{1}{m} \p(U=u)$ for
$0 \leq i \leq m-1$.
\item[(ii)] $\p(V^*=v_i) = \frac{1}{m}
\p(V=v)$ for $0 \leq i \leq m-1$.
\item[(iii)]
$\p(X^*=k|U^*=u_i) = \p(X=k|U=u)$ for $0 \leq k,i \leq
m-1$.
\item[(iv)] $\p(X^*=k|V^*=v_i) = \p(X=k|V=v)$
for $0 \leq k,i \leq m-1$.
\end{itemize}
\end{lemma}

\begin{proof}
Observe that
\begin{align*}
&\p(U^*=u_i)\\
 &\quad= \sum_{v \in \mathcal{V}} \sum_{j=1}^m
\p(U^*=u_i,V^*=v_j) \\
&\quad= \sum_{v \in \mathcal{V}}
\sum_{j=1}^m \frac{1}{m} \p(U=u,V=v,X=(i-j)_m) \\
&\quad = \sum_{v \in \mathcal{V}} \frac{1}{m} \p(U=u,V=v) \\
&\quad = \frac{1}{m} \p(U=u).
\end{align*}

Proof of (ii) follows similarly. To show (iii), consider
\begin{align*}
&\p(X^*=k|U^*=u_i) \\
&\quad= \sum_{v \in \mathcal{V}}
\sum_{j=1}^m \p(X^*=k,V^*=v_j|U^*=u_i) \\
&\quad\stackrel{(a)}{=} \sum_{v \in \mathcal{V}}
\p(X^*=k,V^*=v_{(i-k)_m}|U^*=u_i) \\
&\quad = \sum_{v \in \mathcal{V}} \p(V^*=v_{(i-k)_m}|U^*=u_i) \\
&\quad = \sum_{v \in \mathcal{V}}
\frac{1}{m}\frac{\p(X=k,V=v,U=u)}{{P}(U^*=u_i)} \\
&\quad \stackrel{(b)}{=} \sum_{v \in \mathcal{V}} \p(X=k,V=v|U=u)\\
&\quad= \p(X=k|U=u),
\end{align*}
where $(a)$ follows from the fact that the rest
of the terms are zero by construction and $(b)$ follows from (i)
using the fact that $\p(U^*=u_i) =
\frac{1}{m}\p(U=u)$. The proof of (iv) follows similarly.
\end{proof}

The following  corollary follows from the above lemma, the fact
that $X^*$ is a deterministic function of $(U^*,V^*)$, and the fact
that $(U^*,V^*)\rightarrow X^* \rightarrow (Y^*,Z^*),
(U,V)\rightarrow X \rightarrow (Y,Z)$ form Markov chains with
$p(y^*,z^*|x^*)=p(y,z|x)$. The proofs are straightforward and are
therefore omitted.

\begin{corollary} The following hold:
\label{co:entstar}
\begin{itemize}
\item[(i)] ${P}(X^*=i) = {P}(X=i)$ ~for ~$0 \leq i \leq m-1$.
\item[(ii)] $H(Y^*|U^*) = H(Y|U)$.
\item[(iii)] $H(Z^*|U^*)= H(Z|U)$.
\item[(iv)] $H(Y^*|V^*) = H(Y|V)$.
\item[(v)] $H(Z^*|V^*) = H(Z|V)$.
\item[(vi)]$H(Y^*|U^*,V^*) = H(Y^*|X^*)$
\item[] $~~~~~~~~~~~~~~~~~~= H(Y|X) \leq H(Y|U,V).$
\item[(vii)] $H(Z^*|U^*,V^*) = H(Z^*|X^*)$
\item[]$~~~~~~~~~~~~~~~~~~ = H(Z|X) \leq H(Z|U,V)$.
\end{itemize}

\end{corollary}

We are now ready to prove Lemma~\ref{le:bisuffices}

\begin{myproof}{of Lemma~\ref{le:bisuffices}}
Corollary \ref{co:entstar} implies that
\begin{equation}
\label{eq:foo}
\begin{aligned}
I(U;Y) &= I(U^*;Y^*),\\
I(V;Z) &= I(V^*;Z^*),\\
I(U;Y|V) &\leq I(U^*;Y^*|V^*),\\
I(V;Z|U) &\leq I(V^*;Z^*|U^*),\\
I(X;Y|V) &= I(X^*;Y^*|V^*),\\
I(X;Z|U) &= I(X^*;Z^*|U^*).
\end{aligned}
\end{equation}
Thus $\cC \subset \cC_d$, which completes the proof of Lemma
\ref{le:bisuffices}.
\end{myproof}

Thus the outer bound in Theorem \ref{th:obproper} can be re-expressed as follows.

\begin{lemma}
\label{le:obound2} The set of rate pairs satisfying
\iffalse
Consider the region $\cC$ formed by taking the
union of the following rate pairs over all choices of distributions
${p}(u,v,x) = {p}(u,v){p}(x|u,v)$, where ${p}(x|u,v) = 0 ~\mbox{or}~
1$.
\fi
\begin{equation}
\label{co:obproper2}
\begin{aligned}
  R_1 & \leq I(U;Y), \\
  R_2 & \leq I(V;Z), \\
  R_1 + R_2 & \leq I(U;Y) + I(X;Z|U),\\
  R_1 + R_2 & \leq I(V;Z) + I(X;Y|V), \notag \\
\end{aligned}
\end{equation}
for some distribution ${p}(u,v,x) = {p}(u,v){p}(x|u,v)$, where ${p}(x|u,v)\in \{0,1\}$, constitute an outer bound on the DM broadcast channel with no
common information.
\end{lemma}

\begin{remark} \label{re:unnec} Note that the constraint
${p}(x|u,v)\in \{0,1\}$ while useful for evaluating the region, can be
removed from the definition, since as before, for any $(U,V,X)$ one
can construct random variables $(U^*,V^*,X^*)$ according to equation
\eqref{eq:constr} and by equation \eqref{eq:foo}, the region
$(R_1,R_2)$ evaluated using $(U,V,X)$ is identical to that evaluated
using $(U^*,V^*,X^*)$.
\end{remark}

\subsection{Cardinality bounds on $U$ and $V$}

We now establish bounds on the cardinality of $U$ and $V$.
From Remark \ref{re:unnec}, we know that $p(x|u,v)$ can
be arbitrary.

\begin{fact} \label{fa:cons} Given $p(u),
p(x|u),{p}(v), p(x|v)$, if $p(x)$ is
consistent, i.e.,
\[ \sum_{u \in \mathcal{U}}
{p}(X=x|u){p}(u) = \sum_{v \in \mathcal{V}}
{p}(X=x|v){p}(v) \]
for every $x \in \mathcal{X}$, then
there exist $p(u,v)$ and $p(x|u,v)$ that are
consistent with $p(u), p(x|u),p(v),
p(x|v)$.
\end{fact}

\begin{remark} A canonical way to generate such a joint triple is to
generate $X$ according to $p(x)$ and then generate $U,V$ conditionally
independent of $X$ according to $p(u|x)$ and $p(v|x)$.
\end{remark}

Now for any $U \ra X \ra (Y,Z)$, using standard arguments from
\cite{ahk75}, there exists a $(U^*,X^*)$ with $\| U^* \| \leq \| X \| +
2$, such that $I(U;Y)=I(U^*;Y^*)$ and
$I(X;Z|U)=I(X^*;Z^*|U^*)$. Similarly, there exists a $V^*$ with $\| V^* \|
\leq \| X \| + 2$, such that $I(V;Z)=I(V^*;Z^*)$ and
$I(X;Y|V)=I(X^*;Y^*|V^*)$. From Fact \ref{fa:cons}, it follows that
there exists a triple $(U^*,V^*,X^*)$ consistent with the pairs
$(U^*,X^*)$ and $(V^*,X^*)$. Thus we can assume that $\| U \|
\leq \| X \| + 2,~~ \| V \| \le \| X \| + 2$.

\subsection{Comparison to K\"{o}rner-Marton outer bound}
\label{subse:km}

The outer bound of K\"{o}rner and Marton \cite{mar79} is given by  $\mathcal{O} = \mathcal{O}_y \cap \mathcal{O}_z$, where $\mathcal{O}_y$ is the set of rate pairs $(R_1,R_2)$ satisfying
\begin{align*}
R_1 & \leq  I(X;Y),\\
R_2 & \leq I(V;Z),\\
R_1 + R_2 & \leq  I(V;Z) + I(X;Y|V),
\end{align*}
for some distribution $p(v)p(x|v)$, and $\mathcal{O}_z$ is the set of rate pairs $(R_1,R_2)$ satisfying
\begin{align*}
R_2 & \leq  I(X;Z),\\
R_1 & \leq I(U;Y),\\
R_1 + R_2 & \leq  I(U;Y) + I(X;Z|V),
\end{align*}
for some distribution $p(u)p(x|u)$.

From Lemma \ref{le:obound2}, it is clear that
$\cC \subset \mathcal{O}_y$ and $\cC \subset \mathcal{O}_z$. Hence
\[\cC \subset \mathcal{O} = \mathcal{O}_y \cap \mathcal{O}_z \] and
 $\cC$ is in general contained in the K\"{o}rner-Marton outer
bound. In the following section, we show that the containment is
strict for the {\em binary skew-symmetric} broadcast channel.

\section{Binary Skew-Symmetric Channel}
\label{se:bssc}

\begin{figure}[ht]
\begin{center}
\begin{psfrags}
\psfrag{X}[r]{$X$}
\psfrag{Y}[l]{$Y$}
\psfrag{Z}[l]{$Z$}
\psfrag{p}[b]{$p$}
\psfrag{1-p}[c]{$1-p$}
\psfrag{0}[c]{$0$}
\psfrag{1}[c]{$1$}
\includegraphics[width=0.45\linewidth,angle=0]{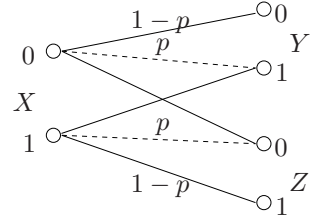}
\end{psfrags}
\caption{Binary Skew
Symmetric Channel}
\label{fig:bssc}
\end{center}
\end{figure}

Consider the Binary Skew-Symmetric Channel (BSSC) shown in Figure
\ref{fig:bssc}, which was studied by Hajek and Pursley
\cite{hap79}. For the rest of the paper we assume that
$p=\frac{1}{2}$, though a similar analysis can be carried out for any
other choice of $p$.

In \cite{hap79}, the Cover-van der Meulen achievable rate region for
the DM  broadcast channel, $\mathcal{D}$, was evaluated for the
Binary Skew-Symmetric broadcast channel (BSSC) with private messages
only. The resulting coding scheme has the following intuitive
interpretation, which we denote by {\em randomized time-sharing}.
Observe that if $X=0$ is sent, it is received error free by $Z$, but
completely noisy by $Y$. Conversely, if $X=1$ is sent, it is
received error free by $Y$, but completely noisy by $Z$. This
suggests that a time-sharing scheme, where transmission time is
divided between the two user before communication commences, is
optimal. It turned out that higher rates can be achieved by
performing randomized time-sharing, instead.  This is done via a
{\em common information} random variable $W$, which specifies the
locations of the symbols in the received sequence corresponding to
each user's {\em private message}. Each receiver first decodes $W$
to find out which part of the received sequence corresponds to its
private message, then proceeds to decode its private message.  Using
standard random coding and joint typicality decoding arguments, it
is can be shown that any $(R_1,R_2)$ satisfying the conditions
\begin{equation*}
\begin{aligned}
  R_1  & \leq \min\{ I(W;Y), I(W;Z)\} \\
  & \qquad+ \p(W=0)I(X;Y|W=0), \\
  R_2 &  \leq \min\{I(W;Y),I(W;Z)\} \\
  & \qquad+ \p(W=1)I(X;Z|W=1),\\
  R_1 + R_2  & \leq \min \{I(W;Y),I(W;Z)\}\\
  &\qquad + \p(W=0)I(X;Y|W=0) \\
  &\qquad + \p(W=1)I(X;Z|W=1),
\end{aligned}
\end{equation*}
for some $p(w)p(x|w)$, is achievable.

The line segment joining $(R_1 ,R_2) = (0.2411..,0.1204..)$ to
$(R_1,R_2)=(0.1204..,0.2411..)$ is achieved by the following
choice of $(W,X)$. Let $\alpha = 0.5 - \sqrt{105}/30\approx
0.1584$, then
\begin{align*}
{\p}(W=0)&= {\p}(W=1)=0.5, \\
{\p}(X=0|W=0) &= \alpha,\\
{\p}(X=1|W=1) &= \alpha.
\end{align*}
It is not difficult to see that the line segment joining $(R_1 ,R_2) =
(0.2411..,0.1204..)$ and $(R_1,R_2)=(0.1204..,0.2411..)$ lies
on the boundary of this region. Note that on this line segment
$R_1 + R_2 = 0.3616...$.

We now show that the region $\cC$, described by the outer bound,
is strictly larger than the Cover-van der Meulen region $\cD$ and
strictly smaller than the K\"{o}rner-Marton outer bound.

\begin{claim} The line segment connecting
$(R_1,R_2)=(0.2280..,0.1431..)$ to $(R_1,R_2)=(0.1431..,0.2280..)$
lies on the boundary of $\cC$.
\end{claim}

\begin{proof}
Note that from Lemma \ref{le:obound2}, the sum rate is bounded by
\begin{equation*}
\begin{aligned}
R_1 + R_2 & \leq \frac{1}{2}(I(U;Y) + I(X;Z|U))\\
&\quad + \frac{1}{2}(I(V;Z) + I(X;Y|V) ).
\end{aligned}
\end{equation*}
We proceed to maximize the RHS of the above inequality over
${p}(u,v,x)$. Assume that $(U_o,V_o,X_o)$ maximizes the sum rate
and let
\begin{equation*}
\begin{aligned}
R_m & = \frac{1}{2}(I(U_o;Y_o) + I(X_o;Z_o|U_o)) \\
&\quad + \frac{1}{2}(I(V_o;Z_o) + I(X_o;Y_o|V_o) ).
\end{aligned}
\end{equation*}
Consider a triple $(U',V',X)$ with $\cU'=\cV, \cV'=\cU$, such that
\begin{equation}
\begin{aligned}
\label{eq:auxrv}
&\p(U'=u',V'=v') \\
&\qquad = \p(U_o=v',V_o=u'),\\
&\p(X=x|U'=u',V'=v')\\
&\qquad = \p(X=1-x|U_o=v', V_o=u').
\end{aligned}
\end{equation}

By the symmetry of the channel,
\begin{align*}
I(U';Y') &= I(V_o;Z_o),\\
I(V';Z') &= I(U_o;Y_o),\\
I(X';Z'|U') &= I(X_o;Y_o|V_o),\\
I(X';Y'|V') &= I(X_o;Z_o|U_o).
\end{align*}
Therefore,
\begin{align*}
R_m &= \frac{1}{2}(I(U';Y') + I(X';Z'|U'))\\
&\quad + \frac{1}{2}(I(V';Z')+ I(X';Y'|V') ).
\end{align*}
Let $Q \in \{1,2\}$ be an independent random variable that takes
values 1 or 2 with equal probability  and define  $U^* =
(\tilde{U},Q)$ and $V^* = (\tilde{V},Q)$ as
$(Q=1,\tilde{U},\tilde{V},X) \sim (U_o,V_o,X)$ and
$(Q=2,\tilde{U},\tilde{V},X) \sim (U',V',X)$, respectively. Then
\begin{equation*}
\label{eq:cvtion}
\begin{aligned}
&\p(X=x|\tilde{U}=u,\tilde{V}=v,Q=1)\\
 &\qquad = \p(X=x|U_o=u,V_o=v),\\
& \p(X=x|\tilde{U}=u,\tilde{V}=v,Q=2) \\
&\qquad= \p(X'=x|U'=u,V'=v).
\end{aligned}
\end{equation*}

Observe that
\begin{equation*}
\label{eq:covxty}
\begin{aligned}
&I(X^*;Y^*|V^*) \\
&\quad = \frac 12 (I(X_o;Y_o|V_o) + I(X';Y'|V')),\quad \qquad \qquad ~~~~\\
\end{aligned}
\end{equation*}
\begin{equation*}
\begin{aligned}
&I(U^*;Y^*) \\
&\quad = H(Y^*) - H(Y^*|U^*) \\
&\quad = H(Y^*) - \frac 12 (H(Y_o|U_o) +
H(Y'|U') \\
&\quad \stackrel{(a)}{\geq} \frac 12 (H(Y_o) + H(Y'))- \frac 12
(H(Y_o|U_o) + H(Y'|U')) \\
&\quad = \frac 12 (I(U_o;Y_o) + I(U';Y')),
\end{aligned}
\end{equation*}
where $(a)$ follows by the concavity of the entropy function.

Similarly
\begin{align*}
I(X^*;Z^*|U^*) &= \frac 12 (I(X_o;Z_o|U_o) + I(X';Z'|U')),\\
I(V^*;Z^*) &\geq \frac 12 (I(V_o;Z_o) + I(V';Z')).
\end{align*}
Therefore,
\begin{equation*}
R_m \le I(U^*;Y) + I(X;Z|U^*) + I(V^*;Z) + I(X;Y|V^*).
\end{equation*}

Now, by the construction of $(U^*,V^*,X^*)$,
${\p}(X^*=1) = 0.5$. Thus to compute $R_m$, it
suffices to consider $X$ such that ${\p}(X=1) = 0.5$.

Using standard optimization techniques, it is not difficult to see
that the following $(U,X)$ and $(V,X)$ maximize the terms  $I(U;Y)
+ I(X;Z|U)$ and $I(V;Z) + I(X;Y|V)$, respectively, subject to
$\p(X=1) = 0.5$. As before, let $\alpha = 0.5 -
\sqrt{105}/30\approx 0.1584$.  Then a set of maximizing pairs
$\p(U,X)$ and $\p(V,X)$ can be described by
\begin{eqnarray*} {\p}(U=0) =
\frac{0.5}{1-\alpha},& &{\p}(U=1) = \frac{0.5 -
\alpha}{1-\alpha},\\ {\p}(X=1|U=0) = \alpha,& &
{\p}(X=1|U=1) = 1, \\ {\p}(V=0) = \frac{0.5}{1-\alpha},&
& {\p}(V=1) = \frac{0.5 - \alpha}{1-\alpha},\\
{\p}(X=0|V=0) = \alpha,& & {\p}(X=0|V=1) = 1.
\end{eqnarray*}
Substituting these values, we obtain
\begin{equation*}
\begin{aligned}
R_1+R_2 & \leq \frac{1}{2}( I(U;Y) + I(X;Z|U) ) \\
& \quad + \leq \frac{1}{2}(I(V;Z) + I(X;Y|V) )\\
& \le 0.3711...
\end{aligned}
\end{equation*}
We now show that this bound on the sum rate is tight.

As before, let $\alpha = 0.5 - \sqrt{105}/30\approx 0.1584$.
Consider the following $(U,V,X)$
\begin{eqnarray*}
{\p}(U=0,V=0) &= &\frac{\alpha}{1-\alpha},\\
{\p}(X=1|U=0,V=0) &= &0.5, \\ {\p}(U=0,V=1) &= &\frac{0.5
- \alpha}{1-\alpha},\\ {\p}(X=1|U=0,V=1) &= &0, \\
{\p}(U=1,V=0) &= &\frac{0.5 - \alpha}{1-\alpha}, \\
{\p}(X=1|U=1,V=0) &= &1.
\end{eqnarray*}

The region evaluated by this $(U,V,X)$ is given by all rate pairs
$(R_1,R_2)$ satisfying
\begin{equation}
\label{eq:imp3a}
\begin{aligned}
  R_1 & \leq I(U;Y) = 0.2280.., \\
  R_2 & \leq I(V;Z) = 0.2280.., \\
  R_1 + R_2 & \leq  I(U;Y) + I(X;Z|U) = 0.3711...\\
  R_1 + R_2 & \leq I(V;Z) +
  I(X;Y|V)  = 0.3711...
\end{aligned}
\end{equation}

Thus the line segment joining $(R_1,R_2)=(0.2280..,0.1431..)$ to
$(R_1,R_2)=(0.1431..,0.2280..)$ lies on the boundary of $\cC$.
\end{proof}

The line segment joining $(R_1,R_2)=(0.2280..,0.1431..)$ to
$(R_1,R_2)=(0.1431..,0.2280..)$ that lies on the boundary of $\cC$
is strictly outside the line segment joining $(R_1 ,R_2) =
(0.2411..,0.1204..)$ to $(R_1,R_2)=(0.1204..,0.2411..)$ that lies on
the boundary of the Cover-van der Meulen region $\mathcal{D}$
\cite{hap79}.

Consider the following random variables $(U,X)$.
\begin{align*}
{\p}(U=0)&=0.6372,\\
{\p}(U=1)&=0.3628,\\
{\p}(X=1|U=0) &= 0.2465,\\
  {\p}(X=1|U=1) &= 1.
\end{align*}
For this pair $I(U;Y) = 0.18616..$ and $I(X;Z|U) = 0.18614..$. Hence
the point $(R_1,R_2) = (0.1861,0.1861)$ lies inside the region
$\cO_y$. By symmetry, the same point lies inside $\cO_z$ and hence it
lies inside $\cO_y \cap \cO_z$, the K\"{o}rner-Marton outer
bound. Note that $R_1 + R_2 = 0.3722> 0.3711.. $ and therefore this point
lies outside $\cC$.

\section{Conclusion}
We presented a new outer bound on the capacity region of the DM
broadcast channel (Theorem 2.1), which is tight for all special cases
where capacity is known. We then specialized the bound to the case of
no common information (see (3.1)).  Considering the weaker version of
this bound given in Theorem 3.1, we showed that our general outer bound is
strictly smaller than the K\"{o}rner-Marton for the BSS channel. The
outer bound in Theorem 3.1, however, is strictly larger than the
Cover-van der Meulen region for this channel.  We suspect that in
general the outer bound in \eqref{eq:obound} is strictly tighter than
that in Theorem \ref{th:obproper}. We have not been able to verify
this for the BSS due to the complexity of evaluating
(3.1). Finally, it would be interesting to show that our new outer
bound is tight for some new class of broadcast channels that may
perhaps include the BSSC.

\bibliographystyle{plain}
%\begin{thebibliography}{10}

%\bibitem {A} T. Aoki, \textit{Calcul exponentiel des op\'erateurs
%microdifferentiels d'ordre infini.} I, Ann. Inst. Fourier (Grenoble)
%\textbf{33} (1983), 227--250.

%\bibitem {B} R. Brown, \textit{On a conjecture of Dirichlet},
%Amer. Math. Soc., Providence, RI, 1993.

%\bibitem {D} R. A. DeVore, \textit{Approximation of functions},
%Proc. Sympos. Appl. Math., vol. 36,
%Amer. Math. Soc., Providence, RI, 1986, pp. 34--56.

%\end{thebibliography}
%\bibliography{mybiblio}

\begin{thebibliography}{10}

\bibitem{ahk75}
R~F Ahlswede and J~K\H{o}rner.
\newblock Source coding with side information and a converse for degraded
  broadcast channels.
\newblock {\em IEEE Trans. Info. Theory}, IT-21(6):629--637, November, 1975.

\bibitem{ber73}
P~F Bergmans.
\newblock Coding theorem for broadcast channels with degraded components.
\newblock {\em IEEE Trans. Info. Theory}, IT-15:197--207, March, 1973.

\bibitem{cov72}
T~Cover.
\newblock Broadcast channels.
\newblock {\em IEEE Trans. Info. Theory}, IT-18:2--14, January, 1972.

\bibitem{cov75}
T~Cover.
\newblock An acheivable rate region for the broadcast channel.
\newblock {\em IEEE Trans. Info. Theory}, IT-21:399--404, July, 1975.

\bibitem{czk78}
I~Csiz\'{a}r and J~K\H{o}rner.
\newblock Broadcast channels with confidential messages.
\newblock {\em IEEE Trans. Info. Theory}, IT-24:339--348, May, 1978.

\bibitem{elg79}
A~El~Gamal.
\newblock The capacity of a class of broadcast channels.
\newblock {\em IEEE Trans. Info. Theory}, IT-25:166--169, March, 1979.

\bibitem{gal74}
R~G Gallager.
\newblock Capacity and coding for degraded broadcast channels.
\newblock {\em Probl. Peredac. Inform.}, 10(3):3--14, 1974.

\bibitem{gep80}
S~I Gelfand and M~S Pinsker.
\newblock Capacity of a broadcast channel with one deterministic component.
\newblock {\em Probl. Inform. Transm.}, 16(1):17--25, Jan. - Mar., 1980.

\bibitem{hap79}
B~Hajek and M~Pursley.
\newblock Evaluation of an achievable rate region for the broadcast channel.
\newblock {\em IEEE Trans. Info. Theory}, IT-25:36--46, January, 1979.

\bibitem{kom75}
J~K\H{o}rner and K~Marton.
\newblock A source network problem involving the comparison of two channels ii.
\newblock {\em Trans. Colloquim Inform. Theory, Keszthely, Hungary}, Auguts,
  1975.

\bibitem{mar77}
K~Marton.
\newblock The capacity region of deterministic broadcast channels.
\newblock {\em Trans. Int. Symp. Inform. Theory}, 1977.

\bibitem{mar79}
K~Marton.
\newblock A coding theorem for the discrete memoryless broadcast channel.
\newblock {\em IEEE Trans. Info. Theory}, IT-25:306--311, May, 1979.

\bibitem{pin78}
M~S Pinsker.
\newblock Capacity of noiseless broadcast channels.
\newblock {\em Probl. Pered. Inform.}, 14(2):28--334, Apr.- Jun., 1978.

\bibitem{sat78}
H~Sato.
\newblock An outer bound to the capacity region of broadcast channels.
\newblock {\em IEEE Trans. Info. Theory}, IT-24:374--377, May, 1978.

\bibitem{van75}
E~van~der Meulen.
\newblock Random coding theorems for the discrete memoryless broadcast channel.
\newblock {\em IEEE Trans. Info. Theory}, IT-21:180--190, March, 1975.

\end{thebibliography}

\newpage

\begin{biographynophoto}{Chandra Nair}
Chandra Nair is a Post-Doctoral researcher with the theory group
at Microsoft Research, Redmond. He obtained his PhD from the
Electrical Engineering Department at Stanford University in June
2005.  He obtained the Bachelor's degree in Electrical Engineering
from IIT, Madras.  His research interests are in discrete
optimization problems arising in Electrical Engineering and
Computer Science, algorithm design, networking and information
theory.  He has received the Stanford and Microsoft Graduate
Fellowships (2000-2004, 2005) for his graduate studies, and he was
awarded the Philips and Siemens(India) Prizes in 1999 for his
undergraduate academic performance.
\end{biographynophoto}

\vspace*{-5.5in}
\begin{biographynophoto}{Abbas El Gamal}
Abbas El Gamal (S'71-M'73-SM'83-F'00) received his B.Sc. degree in
Electrical Engineering from Cairo University in 1972, the M.S. in
Statistics and the PhD in Electrical Engineering from Stanford in
1977 and 1978, respectively. From 1978 to 1980 he was an Assistant
Professor of Electrical Engineering at USC. He has been on the
Stanford faculty since 1981, where he is currently Professor of
Electrical Engineering and the Director of the Information Systems
Laboratory. He was on leave from Stanford from 1984 to 1988 first
as Director of LSI Logic Research Lab, then as cofounder and Chief
Scientist of Actel Corporation. In 1990 he co-founded Silicon
Architects, which was later acquired by Synopsys. His research has
spanned several areas, including information theory, digital
imaging, and integrated circuit design and design automation. He
has authored or coauthored over 150 papers and 25 patents in these
areas.
\end{biographynophoto}

\end{document}